\documentclass[reprint,amsmath,amssymb,prl,superscriptaddress]{revtex4-2}
\usepackage{graphicx}
\usepackage{url}
\usepackage{xcolor}
\usepackage{hyperref}
\usepackage[english]{babel}

\usepackage{letltxmacro}

\LetLtxMacro{\ORIGselectlanguage}{\selectlanguage}
\makeatletter
\DeclareRobustCommand{\selectlanguage}[1]{%
  \@ifundefined{alias@\string#1}
    {\ORIGselectlanguage{#1}}
    {\begingroup\edef\x{\endgroup
       \noexpand\ORIGselectlanguage{\@nameuse{alias@#1}}}\x}%
}
\newcommand{\definelanguagealias}[2]{%
  \@namedef{alias@#1}{#2}%
}
\makeatother

\definelanguagealias{en}{english}

\begin{document}


\title{\papertitle}
\author{Bing Huang}
\email{bing.huang@univie.ac.at}
\affiliation{University of Vienna, Faculty of Physics,  Kolingasse 14-16, AT1090 Wien, Austria}

\author{Guido Falk von Rudorff}
\email{vonrudorff@uni-kassel.de}
\affiliation{University Kassel, Department of Chemistry, Heinrich-Plett-Str.40, 34132 Kassel, Germany}
\affiliation{Center for Interdisciplinary Nanostructure Science and Technology (CINSaT), Heinrich-Plett-Straße 40, 34132 Kassel}

\author{O. Anatole von Lilienfeld}
\email{anatole.vonlilienfeld@utoronto.ca}
\affiliation{Vector Institute for Artificial Intelligence, Toronto, ON, M5S 1M1, Canada}
\affiliation{Departments of Chemistry, Materials Science and Engineering, and Physics, University of Toronto, St. George Campus, Toronto, ON, Canada}
\affiliation{Machine Learning Group, Technische Universit\"at Berlin and Berlin Institute for the Foundations of Learning and Data, 10587 Berlin, Germany}

\title{Towards self-driving laboratories:\\ The central role of density functional theory in the AI age} 

\begin{abstract}
\noindent 120 character summary: 
{\em We review density functional theory's role for efficient, accurate, scalable, and transferable machine learning models.}\\
Density functional theory (DFT) plays a pivotal role for the chemical and materials science due to its relatively high predictive power, applicability, versatility and computational efficiency. 
We review recent progress in machine learning model developments which has relied heavily on density functional theory for synthetic data generation and for the design of model architectures. 
The general relevance  of these developments is placed in some broader context for  the chemical and materials sciences. 
Resulting in DFT based machine learning models with high efficiency, accuracy, scalability, and transferability (EAST), recent progress indicates probable ways for the routine use of successful experimental planning software within self-driving laboratories.  
\end{abstract}

\maketitle

\section{Introduction}

\begin{figure*}
    \centering
        \includegraphics{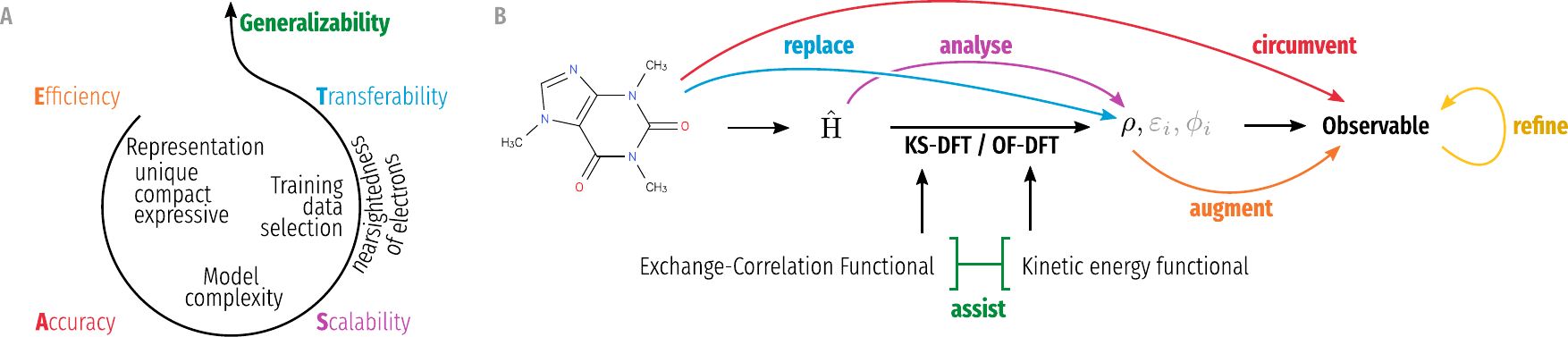}
    \caption{
    Conceptual overview.
    A: To review the key role of DFT, we rely on four key categories of predictive ML models of chemical and materials properties and processes: E (Efficiency), A (Accuracy), S (Scalability) and T (Transferability). A model of EAST generalizes well to unseen systems.
     B: 
   Machine learning approaches going beyond DFT. Black: Traditional workflow in Kohn-Sham (KS) or Orbital Free (OF) DFT where a molecule defines the Hamiltonian which yields the Kohn-Sham-orbitals $\phi_i$ (or electron density, $\rho = \sum |\phi_i|^2$) and eigenvalues $\epsilon_i$. 
   Coloured:  ML based routes taken to build upon DFT.}
    \label{fig:classification}     \label{fig:EAST}

\end{figure*}
It is hardly controversial that we live in an era of artificial intelligence (AI). 
By now AI has touched upon and affected almost any branch of human activity,
 assuming centre stage in many domains of daily life, such as natural language processing, computer vision, or forecasting. 
The dream of an AI based robot scientist (named 'Adam') to assist and accelerate scientific discovery has been introduced to synthetic biology already nearly twenty years ago~\cite{king2004functional}.
Promising first steps with regards to robotic and autonomous experimentation for chemistry and materials, however, have only been made most recently\cite{ley2015organic,granda2018controlling,sanderson2019automation,coley2019robotic,burger2020mobile,vaucher2020automated,park2023closed}, 
e.g.~delivering self-driving laboratories for thin film discoveries~\cite{macleod2020self}).
Nevertheless, such ground-breaking progress has already led Krenn et al~ to survey community members and to fundamentally reconsider the meaning of 
``understanding'' in the context of the scientific process itself~\cite{krenn2022scientific}. 
As already outlined previously by Aspuru-Guzik, Lindh, and Reiher~\cite{aspuru2018matter}, the success of autonomous self-driving labs in chemistry and materials will depend crucially on the availability of machine learning based control software capable to reliably forecast and rank experimental outcomes throughout the materials and chemical compound space (CCS) with sufficient accuracy in real time~\cite{Anatole2020NatureReview}.
CCS refers to the tremendously large set that emerges for all conceivable combinations of chemical composition, structures, and experimental conditions that result in stable forms of matter~\cite{ChemicalSpace}. 
Thermodynamic and kinetic stability being well defined via the quantum statistical mechanics of electrons and nuclei, reliance on a quantum mechanics based approach towards CCS is as obvious as {\em alternativlos}. 
Unfortunately, the relevant equations of quantum and statistical mechanics can only be solved exactly for the simplest of systems, rendering numerical solutions of approximate expressions necessary. 
Method development for the physics based computational design and discovery of materials and molecules in CCS represents a long-standing challenge  and has motivated decades of atomistic simulation research~\cite{ceder1998predicting,marzari2016materials}
Applications are as diverse as the atomistic sciences and include improved solutions for batteries, transistors, catalysts, coatings, ligands, alloys, or photo-voltaics, among others. 
All such efforts have in common that they attempt to virtually navigate CCS in order to narrow down the search space for subsequent experimental verification and characterization.

The importance of electronic structure information for computational materials identification, 
characterization and optimization has recently been highlighted by Marzari, Ferretti, and Wolverton~\cite{marzari2021electronic}
The probably most powerful compromise between predictive power and computational burden for calculating properties and behavior of gaseous and condensed systems from first principles is Density Functional Theory (DFT). 
In particular, the effective single-particle flavour of DFT, approximating the electronic kinetic energy contribution within the Kohn-Sham framework~\cite{KS} has proven immensely useful.
A sheer countless number of ever improving approximations to the exact yet unknown exchange-correlation potential has led to further progress\cite{Perdew2017Science}.
With one of the co-founders of DFT, Walter Kohn, having been awarded the Nobel Prize in Chemistry in 1998, expectations for further improvements ran high at the turn of the century~\cite{Mattsson2002},
and during the wide-spread adaption of {\em ab initio} molecular dynamics~\cite{carparrinello} and time dependent DFT~\cite{Grossbook} to also treat thermal effects and electronic excitations, respectively.
Consequently and unsurprisingly, two DFT related contributions featured among 
the top 10 papers of all times as highlighted in {\em Nature} in 2014~\cite{Noorden2014}.
Ample contemporary reviews have described further improvements~\cite{BurkePerspectives_2012jcp}, 
highlighted the importance of numerical  reproducibility~\cite{lejaeghere2016reproducibility}, 
or emphasized the importance of electron density as a measure of quality, 
in addition to energy~\cite{Medvedev2017} (see Fig.~\ref{fig:tradeoff} (D)).

\begin{figure*}
    \centering
    \includegraphics[width=\textwidth]{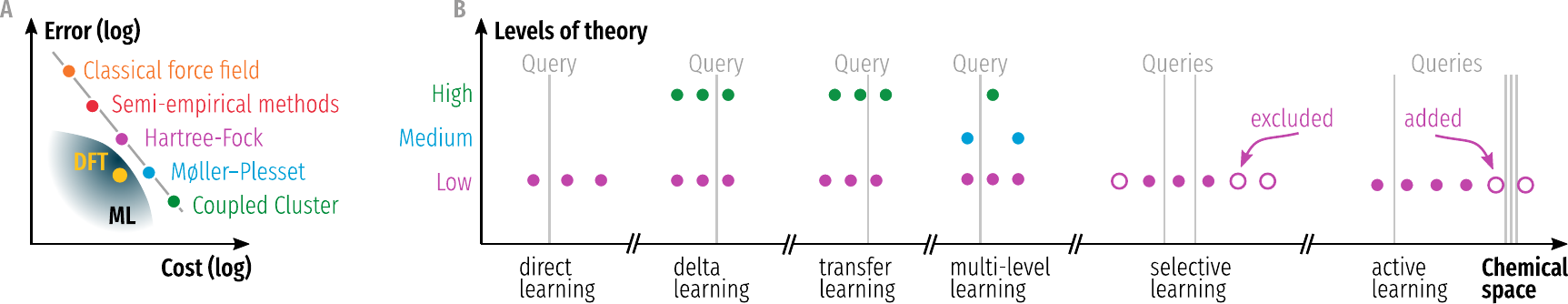}
    \caption{Budget aware compute strategies for sampling chemical compound space. A: Reference data acquisition cost-accuracy Pareto front with popular atomistic simulation approaches. At mean-field cost, DFT stands out approaching the accuracy of explicit electron correlation methods. B: ML model strategies to increase transferability, e.g.~including reference data (points) from different levels of theory for given sets of queries (grey lines). Direct learning refers to all data points from one level, while delta and transfer learning model differences between levels. Multi-level learning accounts for hierarchical data-structures with multiple levels. Selective learning trains on data points (filled) close-by to queries ignoring all others (empty). Active learning biases the data collection (empty) to improve prediction accuracy. }
    \label{fig:pareto}
\end{figure*}

Among all the relevant energy modes that impact the statistical mechanics of matter 
through their partition function (e.g.~translational, rotational, vibrational, etc), 
calculating the electronic structure contribution 
represents the most severe computational bottleneck --- even when using DFT. 
In fact, brute-force execution of DFT based computational chemistry and materials protocols 
consumes major fractions of publicly available high-performance compute allocations~\cite{heinen2020machine}, 
and has still largely failed to become a modern standard of industry 4.0 (`digital twins'). 
This state of affairs is aggravated by (i) CCS's steep combinatorial scaling
and by (ii) neglect of correlations, i.e.~relying on DFT based compute campaigns 
that treat every system independently. 
Encouragingly, relying heavily on DFT physics based supervised 
quantum machine learning (ML) approaches have been introduced over recent years.  
As long as sufficient training data is provided, they have been shown to be universally amenable to the inference of quantum observables while sampling CCS.
The encouraging success of ML in this domain has led to 
many reviews and special issues in the peer-reviewed literature\cite{ceriotti2021editorial}
 and has undoubtedly played an instrumental role in the creation of new journals, such as Springer's {\em Nature Machine Intelligence}, 
IOP's {\em Machine Learning: Science and Technology}~\cite{von2020introducing},
or Wiley's
{\em Applied Artificial Intelligence Letters}~\cite{pyzer2020editorialAIIL}.

We believe that it is difficult to overstate the general importance of these developments. In particular, as also argued previously~\cite{QMLessayAnatole}, the emergence of generalizing statistical surrogate models (machine learning) indicates the formation of a fourth pillar in the hard sciences. This notion is universally applicable, i.e. going even beyond just the chemical and materials sciences. More specifically, first, second, third, and forth pillar respectively correspond to manual experimentation, the theoretical framework to explain and predict experimental observables, numerical simulation tools for computationally complex equations of the theoretical framework, and statistical learning approaches that exploit relations encoded in experimental or simulated training data in order to infer observables. 
These pillars clearly build onto each other, and DFT can be seen as bridging and  encompassing them, all the way from the experimentally observable electron probability distribution via the Hohenberg-Kohn theorems and the Kohn-Sham Ansatz via the many numerical implementations and hardware use cases to its use for training data generation and for informing the design of data-efficient ML model architectures. 
Fig.~\ref{fig:classification} illustrates just some of the possible 
ways that have already been explored to make use of DFT within ML models. 
To substantiate our view towards the key role DFT is playing for the fourth pillar of science, 
we will now review many of the specific ML contributions that have strongly benefited from DFT. 
To this end, this review is structured according to four categories, Efficiency, Accuracy, Scalability, and Transferability (EAST), see Fig.~\ref{fig:EAST}. 
EAST components represent an intuitive ordering principle which allows us to meaningful
discuss, distinguish, and compare some of the most important features necessary 
for building and using digital twins within the chemical and materials sciences.

\section{Efficiency}
\begin{figure*}
    \centering
    \includegraphics[width=\textwidth]{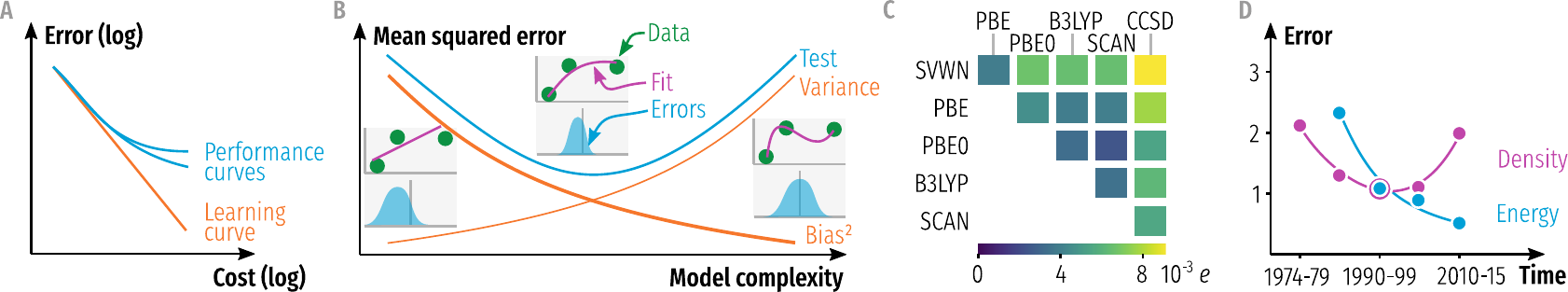}
    \caption{Challenges for machine learning. A: Prediction errors decay with inverse power laws in the limit of large training sets\cite{vapnik1994learningcurves}, while performance curves, i.e. related metrics not explicitly included in loss function, do not necessarily improve arbitrarily. 
    B: Test errors are a combination of a bias from the flexibility of the model to fit the data points and the variance arising from the model's flexibility between data points. C: Integrated DFT based electron density differences, a quantitative measure for density accuracy, of popular density functionals averaged over a random subset of QMrxn20 reactant molecules\cite{Rudorff2020a}. 
    D: DFT errors of energy have improved over time, yet yield lower accuracy for corresponding electron densities (average median-normalised error from literature\cite{Medvedev2017}).}
    \label{fig:tradeoff}
\end{figure*}

Compared to DFT (or higher level quantum chemistry), one of the most striking features of
quantum ML models is their unparalleled prediction speed after training.
While both approaches start from the same information entering the electronic Hamiltonian
in the form of the external potential (i.e.~atomic composition, and geometry), 
ML model predictions are statistical surrogate model evaluations which 
amount to simple and efficient linear algebra operations, typically multiple orders of magnitude faster.
By contrast, conventional physics based simulators, such as
quantum calculations involve the computation of electronic integrals, 
and iterative solvers of differential equations (typically diagonalization),
both of which high-dimensional, non-linear,  and computationally more demanding. 
However, to fully assess the efficiency of ML models, 
the associated cost for both, computational load for testing/training {\em and} 
data-acquisition has to be accounted for. 
Within the context of CCS, data is typically scarce, and the latter point plays a crucial role. 
Training data needs are typically quantified in the form of learning curves, Fig.~\ref{fig:tradeoff}A)
indicating the need for reference data, e.g.~coming from DFT calculations, required to reach a
certain prediction error for out-of-sample queries, i.e.~compounds that have not been part of training. 
{\em Vide infra} for a more detailed discussion of learning curves in the complementary context of the predictive accuracy of QML. 
Given sufficient data, the former point on model complexity becomes the numerical bottleneck. 
ML model complexity is roughly proportional to the number of parameters used for training and testing of models. 
Contrary to parametric neural network or random forest models, non-parametric ML models, such as Gaussian process 
regression, become less efficient as the number of training compounds increases.
Furthermore, increasingly complex representations can reduce model efficiency. 
Considerable contemporary ML research is devoted to maximize the numerical efficiency through hardware (GPUs), as well as model architecture (optimizers, representations)~\cite{browning2022gpu} 

Typically, there is a trade-off between a ML model's efficiency and its predictive power.
For instance, ML models trained on less data can be more efficient but less accurate than those trained on more data.
Sampling training data in representative ways still constitutes a substantial challenge 
when trying to reach universal accuracy already for small fractions of CCS\cite{Rowe2020}.
To better deal with this trade-off between efficiency and capability to generalize,
one can rely on active learning schemes to sample training data more effectively than through random selection.
Active learning (Fig.~\ref{fig:pareto}B) attempts to emulate the optimization of training set selection~\cite{Browning2017}, for example by biasing the training selection using query feature similarity\cite{Smith2018,Huang_2020}. 
One can view this as an attempt to deal with the scaling of compound space\cite{Lookman2019}, 
which is closely related to the idea of `on-the-fly' learning\cite{Csanyi2004}. 
As a result, learning curves are steepened and data efficiency is improved.

Alternatively, one can make use of various flavours of $\Delta$-ML~\cite{DeltaPaper2015,Welborn2018, Bag2021} (\textit{refine} in Fig.~\ref{fig:classification}) 
to enhance predictive accuracy by learning corrections to labels, rather than absolute labels, 
thus exploiting accurately reflected trends in lower level methods as well as error cancellation. 
$\Delta$-ML between DFT and CCSD(T), for example,
has been shown to yield corrected {\em ab initio} molecular dynamics trajectories with improved accuracy\cite{bogojeski2020rhoToDeltaE}.
Similarly, within transfer learning (TL)  optimized neural network weights of the first few layers trained on low-fidelity data (e.g., DFT) can be transferred to the model trained on high-fidelity data (e.g., CCSD)~\cite{Smith2019}.
Though effective in practice, TL may suffer from low explainability. 

\begin{figure*}
    \centering
    \includegraphics[width=\textwidth]{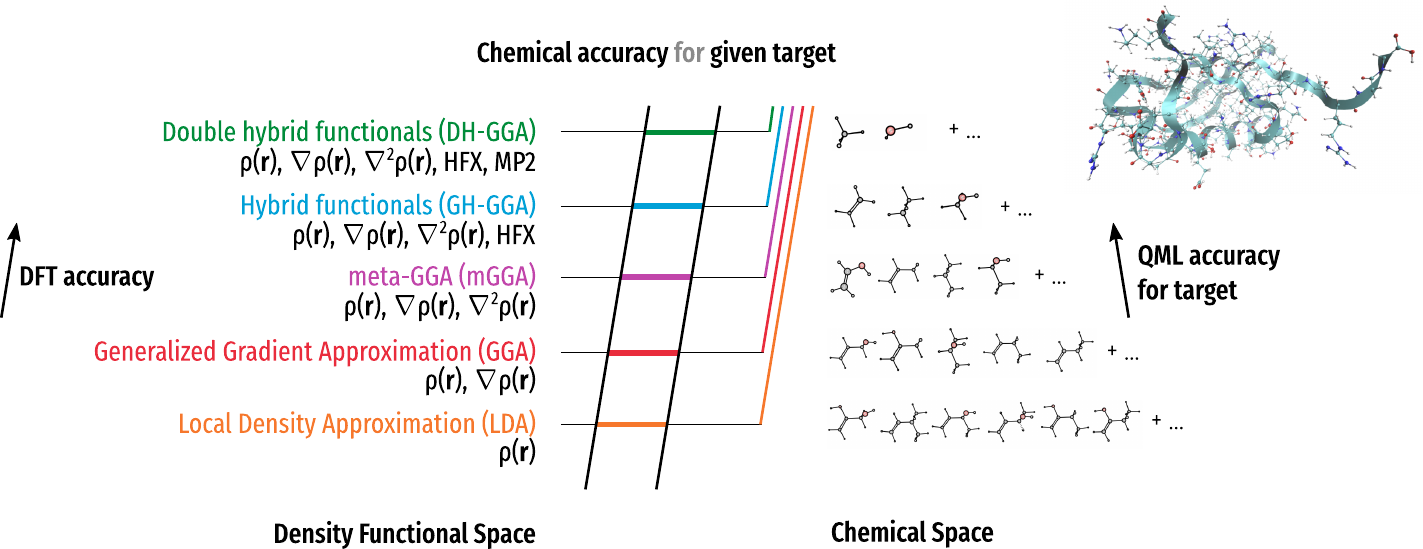}
    \caption{Unifying Jacob's Ladder (Left) in the realm of DFT with amons (Right)~\cite{Huang_2020} space (target-specific hierarchical chemical space) within ML. Predicting properties of large targets (e.g.~small protein ubiquitin shown as inset) with high accuracy while minimizing training data needs. 
    For each pair of amon subset (ranked by system size) and exchange-correlation functional (ranked by Jacob's ladder), the computational burden for training data generation is correlated. First demonstrated for organic chemistry amons and GGAs, hybrid functionals and QMC in Ref.~\cite{deltaAQML2023}
    HFX: Hartree-Fock exchange.
}
    \label{fig:cqml}
\end{figure*}

Alternatively, going beyond just mere $\Delta$-ML~\cite{DeltaPaper2015,deltaAQML2023} and transfer learning~\cite{OutsmartingQuantumChemistry},
correlations between multiple quantum approximations 
can be exploited through use of multi-level grid combinatino (CQML) approaches~\cite{zaspel2018boosting,batra2019multifidelity}. 
CQML is analogous to composite quantum chemistry (often DFT based), 
or Jacob's ladder within DFT (Fig.~\ref{fig:cqml}), 
and allows for systematic error cancellations in hierarchical data sets of basis-set and electron correlation
dimensions, see Fig.~\ref{fig:pareto}B).
For example,  few high-fidelity data (say double-hybrid-DFT) for small systems, 
some medium fidelity data (say PBE0) for medium sized systems,
and many low-fidelity data (say LDA) for large systems can be meaningfully
combined to yield high-fidelity quality predictions (see Fig.~\ref{fig:cqml}).
The CQML model in Ref~\cite{zaspel2018boosting} combines three different dimensions (electron correlation, basis set, training molecules) and levels (strength, size, and number).
Within the realm of DFT, unifying amons~\cite{Huang_2020} (i.e., small molecular fragments
made up of typically no more than 7 non-hydrogen atoms, 
obtained through systematic fragmentation)
and hierarchical density functionals, as demonstrated for the widely-known Jacob's ladder (see Fig.~\ref{fig:cqml}), one
can also greatly improve model efficiency with a set of extra, low-cost calculations (e.g., LDA or GGA).
Consequently, mixing DFT with other high-level electron correlation models (e.g., CCSD(T)/QMC) within the framework of CQML, is a very promising strategy~\cite{deltaAQML2023},
in particular for construction of large-scale data sets of potential energy surfaces (PES). 
In addition to efforts to make DFT property-driven ML more efficient, another promising direction for significant cost reduction is to use ML to help improve orbital-free DFT,
i.e., removing the explicit dependence on orbitals for the kinetic energy (KE) term in KS-DFT by machine-learning the KE density functional\cite{Ryczko2022} directly from data.~\ref{fig:classification}

{\em Perspective} Highly efficient ML models enable us to minimize data needs and model complexity. 
Benefiting in such a way from adherence to the philosophy of Occam's razor, the resulting ML models will enable rapid iterations throughout CCS, virtually as well as in the real world through autonomous robotic execution and evaluation of experiments.

\section{Accuracy}

The underlying promise of all ML is that training on more reference data will result in more accurate models. 
Barring overfitting and inherent limitations due extrapolation, 
any sufficiently flexible regressor converges the prediction error down to numerical noise level in the limit of infinite training data --- consistent with the standard deviation of the error distribution decaying
with the inverse square root of sample number for any fixed domain (central limit theorem). 
The leading term of prediction error typically decays according to an inverse power law for kernel methods as well as neural networks.\cite{vapnik1994learningcurves,StatError_Muller1996}, 
and on log-log scales  \textit{learning curves} (see Fig~\ref{fig:tradeoff}A) must therefore exhibit linearly decaying prediction errors as a function of increasing training set sizes. 
Consequently, learning curves play a crucial role for the comparative assessment of ML models. 
The systematic improvement of a model is only guaranteed for the learned label alone, while derived properties might improve coincidentally, yielding \textit{performance curves} (see Fig.~\ref{fig:tradeoff}A). This bears some analogy to DFT functionals where improvements in accuracy with respect to energies does not imply improvement of electron densities\cite{Medvedev2017} (see Fig.~\ref{fig:tradeoff}D). Note the importance of sufficient flexibility and converged cross-validation to achieve systematic results without overfitting, see Fig.~\ref{fig:tradeoff}B. 

At constant training set size, ML models become more accurate  when hard requirements and boundary conditions are accounted for, e.g.~when including three and four-body interactions~\cite{BAML}
or by reducing delocalisation errors~\cite{Kirkpatrick_2021}.
The representation (feature set)  
that either serves as the ML model input, 
or that is learnt by the ML model itself is crucial. 
As mentioned and reviewed on multiple occasionts~\cite{Anatole2020NatureReview,huang2021abinitio}, training data-efficient 
representations should be unique or complete (in the sense of a bijective one-to-one relationship to the external potential), compact (small size of feature vector), and sufficiently expressive to account for underlying physics such as power-law relations, symmetries, invariances, or constraints. 
A major outstanding challenge is quantifying uncertainty\cite{Reiher2021}, 
especially since  residual errors for common tasks are distinctly non-normal\cite{Pernot_2020}.
Similar ideas have been successful (Jacob's ladder, see Fig.~(\ref{fig:cqml}))within DFT with LDA, PBE, or SCAN satisfying known constraints for DFT functionals\cite{Sun2015} which has helped the accuracy of DFT despite its use of uncontrolled approximations, with only few bounds on the discretisation error available\cite{Cances2021}.

Most DFT based ML models introduced so far roughly fall into any one of three categories (see Fig.~\ref{fig:classification}).
\begin{itemize}
    \item [(a)]
Learning of electronic observables, i.e.~QM averages for which the electronic details of density or wavefunctions have been integrated out (\textit{circumvent/replace} in Fig.~\ref{fig:classification}), relies on input based on the same information as for the model Hamiltonian: atomic coordinates, 
nuclear charges, 
number of electrons, 
and multiplicity. 
Once trained, direct ML is analogous to black-box usage of DFT,
and most common so far.
Its main draw-back is that it may fall short of desired accuracy due to exorbitant training data needs. 
\item[(b)] Recently introduced hybrid ML/DFT approaches improve the DFT model construction. I.e. learning the effective Hamiltonian as an intermediate quantity (\textit{replace} in Fig.~\ref{fig:classification}), from which target properties follow straightforwardly, e.g. SchNorb~\cite{SchNorb} or DeepH~\cite{Li2022DeepH}.
Besides improving accuracy, this hybrid approach can offer better accuracy for intensive properties 
such as HOMO/LUMO energy.
Similarly, ML strategies improved semi-empirical quantum chemistry~\cite{Pavlo2015parameterML} and tight binding DFT~\cite{elstner2018unsupervisedTightBinding,Tkatchenko2020repulsionTightBindingDFT}.
\item[(c)]
Machine-learned density functionals (\textit{assist} in Fig.~\ref{fig:classification}) come in two variants. The first within the KS-DFT framework improving the mapping from the electron density to the exchange-correlation (XC) energy based on higher level reference data (e.g., DFT/CCSD density and CCSD(T) energy).
This way, it is possible to approximate the exact yet unknown XC functional with high accuracy.
The main drawback is that the computational cost of DFT is not alleviated, as typically the explicit dependence on 
orbitals in the kinetic energy and exchange term is not removed.
Well-known contributions include NeuralXC~\cite{Dick2020NeuralXC}) and DM21\cite{Kirkpatrick_2021}.
The second  strategy is more in the spirit of Hohenberg-Kohn building the orbital-free map from electron density to energy (\textit{augment} in Fig.~\ref{fig:classification}).
Ground-breaking contributions include the kinetic energy density functional\cite{Snyder2012MLKErho} and the ML-HK map from potential to density to energy.~\cite{kiat2019MLHK, bai2022MLHK_excitedstat}  
\end{itemize}

{\em Perspective} 
We note that numerical approximations in DFT are rooted in careful neglect of certain physical effects. 
To reach experimental accuracy, however, machine learning models will eventually even require the inclusion of experimental observables in order to guarantee improvement over synthetic computational approaches, e.g. by automated and data-driven approaches\cite{Steiner_2019}. 
Given the strategies outlined, we believe that ML on DFT is reaching a point where such reliable forecasting of materials' properties is possible that autonomous experimental exploration will be necessary to further improve accuracy and applicability.

\section{Scalability}

Scalability is critical for enabling the study of larger and more complex electronic systems.
While linear scaling DFT based implementations such as ONETEP have made great progress~\cite{prentice2020onetep}, 
DFT generally scales cubically with system size which is considerably more favorable than accurate post-HF methods such as CCSD(T) ($\sim\mathcal{O}(N^7)$).
Nevertheless, the routine use of ab initio molecular dynamics simulations using the most accurate flavours of DFT, i.e.~hybrid or range-separated DFT in particular, becomes rapidly elusive for larger systems, say small proteins such as ubiquitin (Fig.~(\ref{fig:cqml}).
Machine learning can also help in this regard, most commonly by partitioning models of extensive properties onto atomic contributions. 
 

Scalability of ML models can also be assessed in a more chemical sense, namely its ability to generalize to larger query systems after training on smaller systems only.
This rests upon the locality assumption which is implied when using similarity measurements that are based on atomic environments. 
This assumption is often justified with reference to Kohn's 
 nearsightedness of electronic matter ~\cite{KohnNearsightedness}.
As pointed out by Kohn and others~\cite{baer_sparsity_1997,kohn_density_1996}, the locality of the one-particle electron density matrix of a system with periodic boundary condition is decays exponentially with the HOMO-LUMO  gap. 
Nearsightedness has a series of consequences, such as the extent of charge transfer (which has bigger impact on charged species), 
conjugation, electron correlation (in particular for strongly correlated systems) and London dispersion (in large biomolecules as well as non-covalently interacting molecular complexes) (see also Fig.~\ref{fig:EAST}).
In practise, a scalable ML model has the necessary but not sufficient requirement of sparsification the representation
such that negligibly strong interactions are omitted. 
The lack of scalability ranks perhaps among the most common and severe challenges of state-of-the-art ML models, possibly due to the the difficulty of accurately including long-range interactions. 
Among the many long range effects, interacting Coulombic multipole moments and their polarizabilities have been considered first within multi-scale ML models~\cite{grisafi2021multi} due to their well known classical structure and the availability of many empirical models.
As for other long range effects of quantum mechanical origin, such as conjugation, spin-orbit-coupling, surface-crossing, and electron correlation in general, robust scalable ML based models are yet to be devised.

{\em Perspective} Once short- and long-range effects, accessible through DFT, are properly accounted for by scalable ML models, the routine study of condensed systems, macro-molecules, defects, and maybe even grain boundaries will become affordable to the entire community.

\section{Transferability}

The physical origins of chemical transferability of atoms, bonds or functional groups 
have been studied much and are so closely related to the aforementioned locality and nearsightedness of electronic matter~\cite{fias2017ChemicalTransferabilityFunctional}
that it is difficult to rigorously separate them.
Here, we associate transferability predominantly with the capability to generalize short-range effects across CCS.
Long-range effects are often less subtle and complex in comparison, and matter more in the context of scalability, as discussed in the previous section.
Examples for short-range effects effects include well-known assignment of interatomic many-body contributions in terms of force-field topologies such as the degrees of freedom associated with covalent bonding, angules, or torsion,
and steric hindrance. 
As illustrated by the amon concept~\cite{Huang_2020} (Fig.~\ref{fig:cqml}) for various labels calculated with DFT, 
these projections of internal degrees of freedom, 
together with their off equilibrium distortions for every conformer, 
can be conveniently represented throughout chemical space by molecular fragments with systematically increasing size.
Other effects, dominated by long-range characteristics, can be left for approximate treatments with empirical
or other relevant models~\cite{Ko2021}).
As a recent example of the insight gained from such a transferable approach, as implemented by training on both, local small bottom-up as well as large top-down fragments, we refer to the stark differences observed between smooth and accurate DFT trained ML trajectories of the protein crambin in aqueous solution, and classical force-field based Brownian motion like counterparts with stochastic characteristics~\cite{unke2022accurate}. 


ML practitioners not familiar with DFT might not be aware of the fact that different quantum properties 
 exhibit strongly varying degrees of transferability, depending on their exact definition. 
The transferability of energy, for instance, depends on whether it is the total energy (derived from a particular theoretical level), an orbital's eigenvalue, an atom's energy within a molecule~\cite{AtomicAPDFT}, the electron correlation energy, reorganization or relaxation energies, or the relative energy  between two theoretical levels.
For example, the latter point was shown recently indicating that ML models of molecular Hartree-Fock (HF) energies require more training data than energies calculated with methods that include electron correlation such as DFT or QMC,
possibly because on average correlation tends to smoothen the label function by contraction of atomic radii~\cite{deltaAQML2023}.
Or, consider the increase in transferability (and consequently reduction in data needs)
for absolute energies vs.~energy differences between two levels of theory as exemplified
for GFMP2 and DFT for $\Delta$-ML already in 2015~\cite{DeltaPaper2015}.

While early studies mostly focused on proof-of-principle and atomic and interatomic energetics throughout CCS, 
more recent research indicates that properties at the finer electronic resolution can be transferable as well.
For instance, transferable ML models of atomic multi-pole moments or NMR shifts calculated with DFT 
were published already in 2015~\cite{MTP_Tristan2015, MLatoms_2015}. 
Or, the electron density of small molecules such as 
ethene 
or butadiene
was found to be transferable to 
octa-tetraene~\cite{Grisafi2018}.
Learning 
the deformation density~\cite{Low2022}, can also help improve the transferability.
Further evidences
include the inter-electronic force field approach~\cite{CoolsCeuppens2022},
and the localized molecular orbital feature based approaches~\cite{Husch2021}.
Superior model transferability can also arise from the use of electronic features like Mulliken charges and bond order~\cite{Duan2022}.

Another important concept at the electronic level, the density functional (DF),
has been deemed transferable at a more fundamental level.
Approximated DFs constructed from (biased) heuristics, however, are prone to severe transferability issues.
Such short-comings, 
also evinced by Nagai and collaborators~\cite{Nagai2020},
could be alleviated through in-depth analysis of prediction error distributions~\cite{pernot2020impact}
leading to more systematic generation of DFs via data-driven ML approaches (\textit{assist} in Fig.~\ref{fig:classification})~\cite{Snyder2012MLKErho}.
By incorporating additional physical constraints into ML, one can enhance the transferability of DFs. 
This has also been shown more recently by learning the non-local exchange DF with a tailored representation that preserves the density distribution under uniform scaling~\cite{Bystrom2022}.
Beyond DFT, physical constraints (such as Kato's cusp conditions) are also crucial in the correlated framework, 
as they help enhance the expressiveness of deep neural network for wavefunction approximation.~\cite{Hermann2020}

{\em Perspective} High transferability across CCS represents the ultimate test to DFT as well as ML. 
Encouraging progress has been made so far, indicating the possibility of sampling CCS more freely; 
concurrently
paving the way towards software control solutions which will routinely deal even with
exotic chemistries and formulations within self-driving lab settings.

\section{Conclusions} 
We have reviewed the instrumental role DFT has been playing for the emergence of machine learning based models
that enable the navigation of chemical compound space with EAST (efficiency, accuracy, scalability, and transferability).
In addition to its fundamental theoretical underpinnings in terms of a quantum mechanical approximation method, 
DFT has served as a truly outstanding source of calculated properties 
for freely chosen molecules or materials 
--- with controllable and reasonable acquisition costs and most welcome accuracy.
Outstanding electronic structure challenges to DFT based ML include surface crossing, open-shell and spin-orbit coupling effects, conductivity and excited states dynamics. 
But even her, DFT can already be extremely useful as an intermediate quality method that can be exploited in multi-fidelity ML models, or for helpful inspiration of physics based ML architectures.  

Availability of large, diverse and high-accuracy (at the experimental level or above) materials and molecular property datasets will remain a fundamental requirement for the development and training of transferable machine learning models that can universally handle any property and chemistry, and that can conveniently be incorporated within the experimental planning software of future self-driving and closed-loop autonomous experimentation.
Conversely, we consider the general scarcity and lack of data to represent the most severe current road-block on our community's path towards that goal.  
For example, many available databases report only select properties and molecular graphs (let alone structures) for Lewis structure conforming systems. 
Notable exceptions, such as DFT based distortions along normal modes and conformers reported in QM7-X~\cite{hoja2021qm7}
or subtle electronic effects such as in carbenes~\cite{QMspin}, 
are few and scarcely scattered throughout the chemical compound space.
Representative data encoding transition states, defects, charged species, radicals, entire ab initio molecular dynamics trajectories, 
$d$- and $f$ elements, or excited states, to name just some, are still mostly lacking.

While notable achievements have been made in successfully applying physics based machine learning to DFT solutions throughout the chemical and materials sciences, there remains a dearth of theoretical research focused on the underlying fundamentals.
Some of the basic open questions  include, 
(i) can one rigorously define CCS in a mathematical way, akin to the Hilbert space for electronic wave-functions, in order to quantify its inherent properties such as density and volume? 
(ii) CCS is discrete in reality, yet which maps enable smooth interpolations into latent spaces that facilitate inverse design~\cite{gomez2018automatic}?
iii) can there be a single ML model that allows for a unified yet accurate description of any chemical compound, regardless of its size, composition, aggregation state, and external conditions?
In summary and roughly speaking, DFT has impacted ML 
as an \textit{ab initio} solver, with ever-improving performance of ML based exchange-correlation and/or kinetic energy density functionals, 
as a hybrid DFT/ML framework for building effective Hamiltonians,
and as a robust computational work-horse for generating highly relevant and affordable synthetic data.
Reflecting on the four EAST categories: Efficiency, accuracy, scalability, and transferability, 
we think it is clear that DFT has played a pivotal role for all the chemical sciences 
in bridging all the pillars of modern science, from
experiments via theory and simulation to physics based  ML model building. 
Seeing the impressive progress made by building on DFT, 
we do not think that it is far-fetched to expect these developments to directly lead to the 
development and wide-spread adaptation of autonomous experimentation. 
Consequently, and building on the four preceding pillars, 
we might very well witness the  emergence of the next, 5th pillar of science 
in a not too distant future: Self-driving laboratories throughout the hard sciences! 

\section{Acknowledgments}
O.A.v.L. has received support from the European Research Council (ERC) 
under the European Union’s Horizon 2020 research and innovation programme (grant agreement No. 772834), 
and as the Ed Clark Chair of Advanced Materials and Canada CIFAR AI Chair.

\bibliography{nablachem,literatur,new}
\end{document}